\documentclass[sigconf,natbib,authorversion]{acmart}

\usepackage{balance}
\usepackage[inline]{enumitem}
\usepackage{stfloats}
\usepackage{amsfonts}
\usepackage[T1]{fontenc}
\usepackage[utf8]{inputenc}
\usepackage{makecell}
\usepackage{multirow}
\usepackage{subcaption}

\usepackage[english]{babel}
\usepackage{numprint}
\usepackage{latexsym}
\usepackage[htt]{hyphenat}

\usepackage{listings}
\everypar{\looseness=-1}
\newcommand{\RQRef}[1]{\textbf{RQ#1}}

\npdecimalsign{.}

\copyrightyear{2024}
\acmYear{2024}
\setcopyright{rightsretained}
\acmConference[SIGIR '24]{Proceedings of the 47th International ACM SIGIR Conference on Research and Development in Information Retrieval}{July 14--18, 2024}{Washington, DC, USA}
\acmBooktitle{Proceedings of the 47th International ACM SIGIR Conference on Research and Development in Information Retrieval (SIGIR '24), July 14--18, 2024, Washington, DC, USA}

\acmDOI{10.1145/3626772.3661355}
\acmISBN{979-8-4007-0431-4/24/07}

\makeatletter
\gdef\@copyrightpermission{
 \begin{minipage}{0.3\columnwidth}
  \href{https://creativecommons.org/licenses/by-nc-nd/4.0/}{\includegraphics[width=0.90\textwidth]{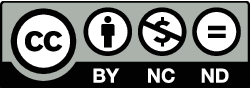}}
 \end{minipage}\hfill
 \begin{minipage}{0.7\columnwidth}
  \href{https://creativecommons.org/licenses/by-nc-nd/4.0/}{This work is licensed under a Creative Commons Attribution-NonCommercial-NoDerivs International 4.0 License.}
 \end{minipage}
 \vspace{5pt}
}
\makeatother

\begin{document}

\title[Synthetic Query Generation using Large Language Models for Virtual Assistants]{Synthetic Query Generation using Large Language Models\\for Virtual Assistants}

\author{Sonal Sannigrahi}
\authornote{Work performed while an intern at Apple.}
\authornote{Equal contribution.}
\email{sonal.sannigrahi@tecnico.ulisboa.pt}
\affiliation{%
  \institution{Instituto Superior Técnico}
  \city{Lisbon}
  \country{Portugal}
}

\author{Thiago Fraga-Silva}
\email{tfragadasilva@apple.com}
\affiliation{%
  \institution{Apple}
  \city{Aachen}
  \country{Germany}
}

\author{Youssef Oualil}
\email{youalil@apple.com}
\affiliation{%
  \institution{Apple}
  \city{Aachen}
  \country{Germany}
}

\author{Christophe Van Gysel}
\authornotemark[2]
\email{cvangysel@apple.com}
\affiliation{%
  \institution{Apple}
  \city{Cambridge, MA}
  \country{USA}
}



\begin{abstract}
Virtual Assistants (VAs) are important Information Retrieval platforms that help users accomplish various tasks through spoken commands. %
The speech recognition system (speech-to-text) uses query priors, trained solely on text, to distinguish between phonetically confusing alternatives. %
Hence, the generation of synthetic queries that are similar to existing VA usage can greatly improve upon the VA's abilities---especially for use-cases that do not (yet) occur in paired audio/text data.
In this paper, we provide a preliminary exploration of the use of Large Language Models (LLMs) to generate synthetic queries that are complementary to template-based methods. We investigate whether the methods (a) generate queries that are similar to %
%
% WARNING: the following language was requested by privacy legal -- do not modify.
randomly sampled, representative, and anonymized %
user queries from a popular VA, and %
(b) whether the generated queries are specific.
We find that LLMs generate more verbose queries, compared to template-based methods, and reference aspects specific to the entity. The generated queries are similar to VA user queries, and are specific enough to retrieve the relevant entity. We conclude that queries generated by LLMs and templates are complementary.

\end{abstract}


\maketitle

\section{Introduction}
\label{sec:introduction}

Virtual Assistants (VAs) are important \cite{Juniper2019popularity} Information Retrieval (IR) platforms that help users accomplish various tasks. Users primarily interact with VAs through voice commands, where users initiate a retrieval request by uttering a query.

The Automated Speech Recognition (ASR) component of the VA system transcribes the spoken user query, which is then subsequently processed by the retrieval engine. However, the ASR system is trained on audio/text pairs that are expensive and time-consuming to obtain. During the recognition process, the ASR system employs a query prior trained solely on text to disambiguate between phonetically-similar recognition candidates. Hence, the query prior is a powerful mechanism to modify the ASR system's behavior, and has been shown to be an effective manner to improve the recognition of tail named entities \cite{Zhang2023server,Saebi2021discriminative,Gondala2021error,VanGysel2020entitypopularity,Pusateri2019interpolation}.

In order to correctly recognize emerging entities \cite{Graus2018birth}, the ASR system's query prior is estimated using a mixture of usage-based and synthetic text data. Synthetic queries are typically generated using a template-based approach \cite{Gandhe2018lmadaptation,VanGysel2022phirtn}. A query template, such as \emph{``play music by \$ARTIST''}, representing the generic intent of a user wanting to play music by a specific artist, is instantiated using a popularity-weighted list of entities. However, template-based approaches are stringent, may only represent a limited set of use-cases, and are not well-suited to generate synthetic queries for use-cases that are specific to particular entities. For example, the query \emph{``play Taylor Swift's debut performance at the Grammy's''} represents the user's intent to play the song \emph{``Fifteen''} by Taylor Swift which was Swift's debut performance at the Grammy's in 2009. While creating a template based on this query would be possible, it does not generalize across entities: some entities may not have performed at the Grammy's and finding the relevant venue would require manual curation. Hence, synthetic query generation methods that can generate queries tailored to specific entities are necessary.  

Recent advances in Large Language Models (LLM) have shown impressive improvements in language understanding tasks \cite{kaplan2020scaling} through their emergent capabilities \cite{Wei2022emergent}. In IR, there have been various works focusing on the generation of queries using LLMs \cite{Alaofi2023llmqueryvariants,Wang2023booleanqueries,Su2023corpussynthesis}.

In this paper, we perform a preliminary analysis of the use of LLMs to produce query priors in VA ASR.
We generate synthetic queries by prompting LLMs using a description of the artist gathered from Wikipedia. Then, we evaluate the generated queries in terms of their \emph{similarity} to %
%
% WARNING: the following language was requested by privacy legal -- do not modify.
randomly sampled, representative, and anonymized %
user queries from a popular VA, in addition to the queries' ability to retrieve the entity for which they were generated.
More specifically, the research questions addressed are as follows: %
\begin{enumerate*}[label=(\textbf{RQ\arabic*})]
    \item Can LLMs generate VA queries that are similar to user queries extracted from VA query logs (i.e., domain match)?
    \item Are the LLM-generated queries good at retrieving the entity for which they were generated (i.e., specificity)?
\end{enumerate*}

Our contributions are as follows: %
\begin{enumerate*}[label=(\arabic*)]
    \item We propose a prompt for LLMs to produce natural queries for the music domain for VAs, and perform extensive experiments comparing the LLM-generated queries to queries generated using template-based methods,
    \item We provide insights through analysis into the differences between queries generated using the various methods.
\end{enumerate*}

\begin{figure*}[t]
    \centering
    \includegraphics[width=\textwidth]{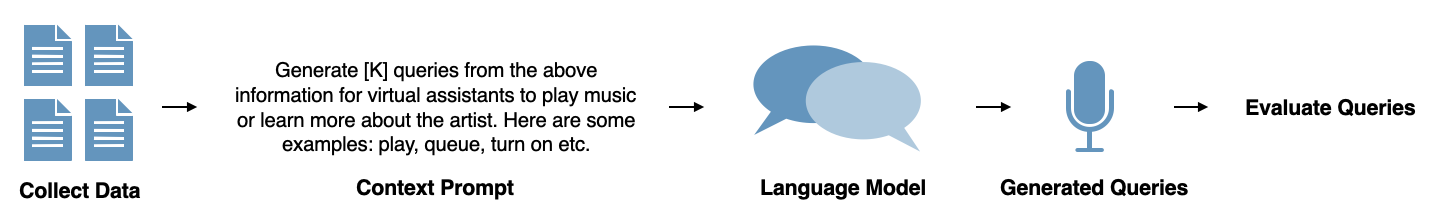}
    \caption{Proposed pipeline to generate queries for a VA via an LLM.\label{fig:pipeline}}
\end{figure*}

\section{Methodology}

Fig.~\ref{fig:pipeline} shows an overview of our approach, which consists of the following three main components: %
\begin{enumerate*}[label=(\arabic*)]
    \item entity descriptions extracted from Wikipedia to provide context for synthetic query generation,
    \item the prompt which incorporates the entity description and formulates a request to the LLM to generate synthetic data, where we also specify the intent the queries need to represent, and
    \item the LLM, which takes the prompt as input and subsequently generates a list of synthetic queries as output.
\end{enumerate*}

\subsection{Knowledge Base}
\label{sec:methodology:kb}

We build our music artist knowledge base by linking Wikipedia data with artist profiles on a popular streaming service. The paragraphs in the Wikipedia articles are used as contexts to generate synthetic queries using LLMs (\S\ref{sec:methodology:prompt}). The combination of the Wikipedia article, and the artist profile retrieved from the streaming service, are used to build a search engine to evaluate the end-to-end performance of the generated queries (\S\ref{sec:results:retrieval}).
We obtained a catalog of music artist entities by %
%
% WARNING: the following language was requested by privacy legal -- do not modify.
downloading %
the list of most popular artists on a streaming service in July 2023 and linking them to their respective Wikipedia profile using property P2850 (i.e., \emph{``Artist ID Number''}) through Wikidata's SPARQL query service\footnote{\url{https://www.wikidata.org/wiki/Wikidata:SPARQL_query_service}}. %
We also use the \emph{MusicGroup}\footnote{\url{https://schema.org/MusicGroup}} metadata object, embedded in the source of each artist page on \url{music.apple.com}, with entries that include artist name, biography, an unique artist ID, as well as discography information. Between both the Wikipedia dumps and the artist database, we maintain a direct linking to produce a knowledge base of \numprint{14161} artists.

\begin{figure}

\begin{lstlisting}[basicstyle=\linespread{0.65}\ttfamily\small, breaklines=true, breakautoindent=false, breakindent=0pt]
[ARTIST DESCRIPTION]

Generate [K] queries based on the information above about [ARTIST NAME] to play music or learn more about [ARTIST NAME].

Here are some examples: [EXAMPLES]
\end{lstlisting}

\caption{LLM prompt used during our experiments. \texttt{[ARTIST DESCRIPTION]} and \texttt{[ARTIST NAME]} are replaced with an entity description, and the entity name, resp. \texttt{[EXAMPLES]} are a list of example VA queries for the specific intent. We fix \texttt{[EXAMPLES]} to the following "play, queue, turn on, etc".  \texttt{[K]} is the number of queries.\label{fig:prompt}}

\end{figure}

\subsection{Prompt \& LLMs}
\label{sec:methodology:prompt}

Our prompt is depicted in Fig.~\ref{fig:prompt}. %
For each entity in the knowledge base (\S\ref{sec:methodology:kb}), we create prompts by populating the artist name and use the lead section (i.e., the introduction) as their description. For music artists, the lead section typically references notable audiography, collaborations and life events. The number of queries, $K$, is set to $40$ (\S\ref{sec:setup:methods}) in this paper.
We use the OpenAI API to generate queries with four model variants\footnote{\url{https://platform.openai.com/docs/models/overview}}. More specifically, we experiment with \texttt{babbage-002}, \texttt{gpt-3.5-turbo}, \texttt{gpt-3.5-turbo-instruct}, and \texttt{gpt-4} (see \S\ref{sec:setup:methods} for details).

\section{Experimental Setup}

\subsection{Query generation methods under comparison}
\label{sec:setup:methods}

We generate queries for all \numprint{14161} entities from our knowledge base (\S\ref{sec:methodology:kb}) using %
\begin{enumerate*}[label=(\alph*)]
	\item the entity name by itself,
	\item a template-based approach using the top-$K$ (according to prior probability) music query templates released as part of \citep{VanGysel2022phirtn} (excluding the templates that consist of only the entity name by itself in order to differentiate from approach (a)), and %
	\item four different LLMs available via OpenAI's API using the prompt in Fig.~\ref{fig:prompt} (\S\ref{sec:methodology:prompt}) where we ask the LLM to generate $K$ queries: \texttt{babbage-002}, \texttt{gpt-3.5-turbo} (v0613), \texttt{gpt-3.5-turbo-instruct}, and \texttt{gpt-4}; with $K = 40$. During our experiments, we report evaluation measures at various values of $K \leq 40$, in which case we extract the first $K$ queries from the list of 40 queries (rather than issuing multiple requests to the LLM with varying $K$).
\end{enumerate*}
Generated queries that start with a VA wakeword (e.g., \emph{``hey VA''} where \emph{VA} refers to the name of the assistant), have the prefix corresponding to the wakeword removed. For example, a query \emph{``hey VA play Moderat''} is normalized to \emph{``play Moderat''}. This step aims at avoiding biases towards methods that frequently generate the wakeword during domain match evaluation (\S\ref{sec:results:domain_match}).

\subsection{Evaluation measures}
\label{sec:setup:evaluation}

To answer \RQRef{1}, we measure the likelihood of the generated queries under a 4-gram back-off language model~\cite{Katz1987backoff} estimated on randomly sampled and anonymized user queries over a span of 2 years from a popular VA. We apply Good Turing smoothing, and N-grams that occur infrequently (less than 3 times) in the data are filtered out. The negative log-likelihood (NLL) of a single query $q$ with $\left|q\right|$ terms is defined as %
$
\text{NLL}\left(q\right) = - \left( \sum_{i=1}^{\left|q\right|} \log P\left( q_i \mid q_1 \ldots q_{i - 1} \right) \right)
$,
where $P\left( q_i \mid q_1 \ldots q_{i - 1} \right)$ represents the probability of the term $q_i$ under the 4-gram LM. Using a 4-gram LM, rather than looking for exact matches in query logs, provides us with a flexible approach to score the likelihood of a query, while also having the ability to assign a score to queries not present in the query logs. The lower the NLL, the more a query is likely under VA user query behavior. We report median NLL across a query set of the first $K$ queries for each entity.

For \RQRef{2}, we measure to what capacity the generated queries can retrieve the entity for which they were generated in order to measure query specificity. %
We build an index of our knowledge base \numprint{14161} entities where each entity is represented by its Wikipedia page and its profile on a music streaming service (\S\ref{sec:methodology:kb}), including biography and most popular discography. Both indexed documents and queries are pre-processed by lower-casing, removing punctuation and non-alphanumeric characters, removing stopwords, and applying a Porter stemmer. %
We use the BM25-L retrieval model \cite[\S3.2]{Trotman2014bm} with $k_1 = 1.5$, $b = 0.75$ and $\delta = 0.5$. Since for each query, there is only one relevant entity, we report reciprocal rank (RR), averaged over the top-$K$ queries $q$ and entities $e$, with RR defined as
\begin{equation*}
\text{RR}\left(q, e\right) = \frac{1}{\text{rank}\left(q, e\right)}
\end{equation*}
where $\text{rank}\left(q, e\right)$ equals the rank of the entity $e$ for which query $q$ was generated under the BM25-L retrieval model. The higher the RR, the better a query is able to retrieve the entity it was generated for (and, hence, the more specific an entity is). We report mean RR across a query set for the first $K$ queries generated for each entity.

\section{Results}

\begin{table*}[ht!]
\begin{tabular}{@{}lrrrrrr@{}}%
\toprule%
\multirow{1}{*}{}&\multicolumn{1}{c}{\textbf{entity name}}&\multicolumn{1}{c}{\textbf{templates}}&\multicolumn{1}{c}{\textbf{babbage{-}002}}&\multicolumn{1}{c}{\textbf{gpt{-}3.5{-}turbo}}&\multicolumn{1}{c}{\textbf{gpt{-}3.5{-}turbo{-}instruct}}&\multicolumn{1}{c}{\textbf{gpt{-}4}}\\%
\midrule%
\# entities&14161&14161&13848&14161&14156&14161\\%
\# unique queries per entity&$1.00 \pm 0.00$&$39.51 \pm 0.50$&$9.64 \pm 14.74$&$41.60 \pm 3.61$&$40.09 \pm 2.27$&$39.99 \pm 0.13$\\%
query length per entity&$1.58 \pm 0.80$&$3.78 \pm 1.24$&$52.30 \pm 122.05$&$8.11 \pm 2.66$&$8.42 \pm 4.03$&$8.31 \pm 2.47$\\%
\% of queries with > 15 terms&0.01\%&0.01\%&40.85\%&1.17\%&1.44\%&1.16\%\\\bottomrule%
\end{tabular}
\caption{Statistics of generated queries across the approaches under consideration (\S\ref{sec:setup:methods}).\label{tbl:results:stats}}
\end{table*}

\begin{figure*}[ht!]
\captionsetup[subfigure]{aboveskip=2pt,belowskip=2pt}
\centering
\begin{subfigure}[b]{\textwidth}
\centering
\includegraphics[width=0.80\textwidth]{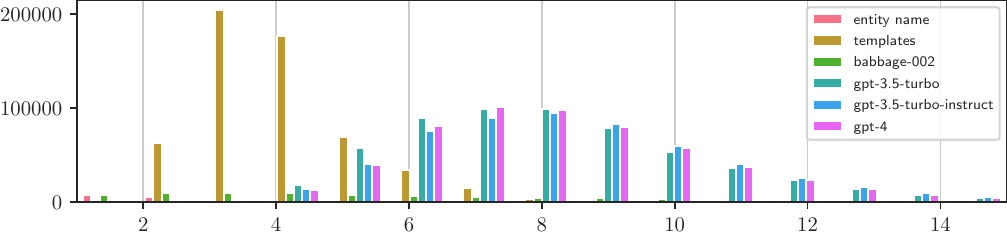}
\caption{Distribution of generated query lengths across the approaches under consideration (\S\ref{sec:setup:methods}). Lengths that exceed 15 tokens are not depicted, but are documented in Table~\ref{tbl:results:stats}.\label{fig:results:query_lengths}}
\end{subfigure}%
\\
\begin{subfigure}[b]{.45\textwidth}
\centering
\includegraphics[width=0.85\columnwidth]{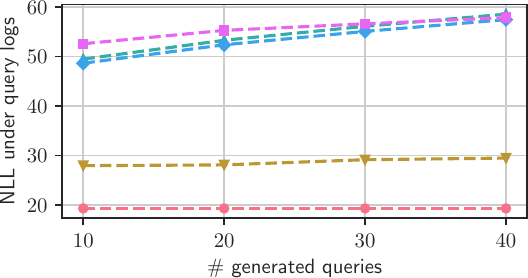}
\caption{Median NLL (\S\ref{sec:setup:evaluation}; lower is better) for the various query generation methods (except \texttt{babbage-002} since it leads to very high NLL) for various query cut-offs ($K = 10, 20, 30, 40$). See Fig.~\ref{fig:results:query_lengths} for the legend.\label{fig:results:nll}}
\end{subfigure}%
\hfill%
\begin{subfigure}[b]{.45\textwidth}
\centering
\includegraphics[width=0.90\columnwidth]{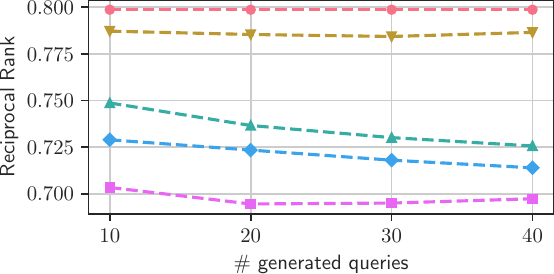}
\caption{Reciprocal rank (\S\ref{sec:setup:evaluation}; higher is better) for the various query generation methods (except \texttt{babbage-002} since it generates non-sensical queries) for various query cut-offs ($K = 10, 20, 30, 40$). See Fig.~\ref{fig:results:query_lengths} for the legend.\label{fig:results:rr}}
\end{subfigure}
\caption{}
\end{figure*}

\begin{table}[ht!]
\resizebox{\columnwidth}{!}{%
\renewcommand{\arraystretch}{0.0}%
\setlength\tabcolsep{1pt}%
\setlength\extrarowheight{-3pt}%
\scriptsize%
\begin{tabular}{@{}ll@{}}%
\toprule%
\textbf{Method} & \textbf{Sample of generated queries} \\
\midrule%
\texttt{\scriptsize entity name} & \emph{``Post Malone''} \\
\midrule%
\texttt{\scriptsize templates} & \makecell[l]{\emph{``play Post Malone''}, \emph{``play the song Post Malone''},\\\emph{``play Post Malone music''}} \\
\midrule%
\texttt{\scriptsize babbage-002} & \emph{``Start with CTRL + M''}, $\ldots$ \\
\midrule%
\texttt{\scriptsize gpt-3.5} & \makecell[l]{\emph{``play White Iverson by Post Malone''},\\\emph{``queue Congratulations by Post Malone''},\\\emph{``turn on Post Malone's album Beerbongs \& Bentleys''}} \\
\midrule%
\makecell[l]{\texttt{\scriptsize{gpt-3.5}}\\\texttt{\scriptsize{(instruct)}}} & \makecell[l]{\emph{``play Post Malone's debut single White Iverson''},\\\emph{``play Post Malone's hit song Rockstar''},\\\emph{``play Post Malone's song Sunflower from the}\\\emph{Spider-Man Into the Spider-Verse soundtrack''}} \\
\midrule%
\texttt{\scriptsize gpt-4} & \makecell[l]{\emph{``play White Iverson by Post Malone''},\\\emph{``add Rockstar by Post Malone to my playlist''},\\\emph{``turn up the volume for Psycho by Post Malone''}} \\
\bottomrule%
\end{tabular}}
\caption{Example of queries generated by the various methods (\S\ref{sec:setup:methods}).
\label{tbl:results:example_queries}}
\end{table}

Table~\ref{tbl:results:stats} shows statistics on the generated queries using the various methods (\S\ref{sec:setup:methods}). In Fig.~\ref{fig:results:query_lengths}, we see that while the standalone entity-name and template-based methods generate relatively short queries (1--4 terms), the LLM-based methods tend to be more verbose ($\sim$8 terms). %
A sample of generated queries for entity \emph{Post Malone} (Q21621919) is depicted in Table~\ref{tbl:results:example_queries}. %
The \texttt{babbage-002} LLM, a GPT base model not trained with instruction following \cite{Ouyang2022instructionft}, performs poorly and fails to generate reasonable queries.
As expected, the template-based approach generates queries that are stylistically simple, since the template is independent from the entity for which the queries are being generated. On the other hand, queries generated by LLM-based methods are able to refer to information present in the artist description that was fed as context to the LLM. %
We will now answer the research questions raised in \S\ref{sec:introduction} and further defined in \S\ref{sec:setup:evaluation}.

\subsection{Similarity to VA usage queries}
\label{sec:results:domain_match}

For \RQRef{1}, Fig.~\ref{fig:results:nll} shows the negative log likelihood (\S\ref{sec:setup:methods}) for the methods under consideration. The \texttt{entity name} by itself aligns closest with user behavior, while the \texttt{template}-based approach is a close second. This is not surprising, since the templates we used were created by domain experts by analyzing high-frequency use-cases in a representative sample of VA usage \cite[\S3.1]{VanGysel2022phirtn}. Hence, the \texttt{entity name} and \texttt{template} method represent frequent use-cases at the head of the query distribution.

Next up, at approximately half the log-likelihood, queries generated by the LLMs seem to represent infrequent, tail use-cases. While not entirely absent from VA usage, they are not as common as the straight-forward templates. This is explained by the fact that the LLM-generated queries often reference specific songs or albums by the artist---extracted from the artist's description---resulting in less generic queries. However, this lack of generality yields queries that reference multiple entities and, hence, tend to be at most as---and often, significantly less---likely as queries referencing only a single entity. Note that in our prompt (Fig.~\ref{fig:prompt}), we did not instruct the LLMs to exhibit this behavior.
We answer \RQRef{1} as follows: queries generated by LLMs trained with instruction following correlate with VA user behavior, although they tend to be more specific than queries generated using template-based approaches. This raises the question whether template- and LLM-based approaches are complementary when it comes to synthetic query generation. In fact, comparing the query sets generated by the \texttt{template}-based method and \texttt{gpt-3.5-turbo-instruct}, the mean/std. dev of the Jaccard coefficient across entities equals $0.0038 \pm 0.0084$, indicating very low overlap, and hence, complementarity.

\subsection{Query specificity}
\label{sec:results:retrieval}

For \RQRef{2}, Fig.~\ref{fig:results:rr} depicts the reciprocal rank for the various methods (\S\ref{sec:setup:methods}) at various cut-offs of the list of generated queries. The \texttt{entity name} method performs best, since it does not contain any superfluous terms and matches directly the specific entity mention contained within the entity's textual representation. The \texttt{template}-based method performs nearly as well as the entity name method, since it generates queries that contain the entity name padded with carrier terms that are non-specific to the entity (e.g., \emph{``play''}, \emph{``song''}).
The LLM-based methods score slightly worse than the \texttt{entiy name} and \texttt{template} methods, since the queries generated using LLMs are more verbose and can include terms that match non-relevant entities. For example, song titles often consist of generic, non-specific terms and multiple songs can have the same title. Between the LLM-based generated query collections, \texttt{gpt-4} performs worst.
When examining the effect of query cut-off ($K$), we see that as $K$ increases, RR generally decreases. This is due to the fact that, as $K$ increases, queries become more complex and can contain terms that confuse the retrieval model.
We answer \RQRef{2} as follows: entity-centric queries generated by LLMs achieve an average reciprocal rank of $0.70$; indicating that the correct entity is often ranked highly. However, since LLMs generate more verbose queries, there are more terms in the queries that can throw off the retrieval model.

\subsection{Complementarity}

Finally, following the conclusions to \RQRef{1} and \RQRef{2} above, and the qualitative examples in Table~\ref{tbl:results:example_queries}, we find that template- and LLM-based methods are complementary as follows: %
\begin{enumerate*}[label=(\arabic*)]
    \item template-based methods allow to generate synthetic queries for frequent use-cases (e.g., for tail entities) that apply across all entities (e.g., \emph{``play music by \$ARTIST''}) and are computationally inexpensive, whereas
    \item LLM-based methods can generate specialized/infrequent use-cases (e.g., for popular/controversial entities) specific to the entity in question (e.g., \emph{``play Taylor Swift's duet with Ed Sheeran''})---while having a higher computational cost.
\end{enumerate*}
Hence, template- and LLM-based methods can be combined to build a richer synthetic query collection with coverage for both (a) tail entities, and (b) tail use-cases.

\section{Conclusions}

In this paper, we performed a preliminary analysis of the use of LLM-based approaches for the generation of synthetic queries for training a query prior used within a VA speech recognition system. We find that template- and LLM-based approaches are complementary since (a) template-based methods can generate queries for frequent use-cases and infrequent entities, and (b) LLM-based methods are better suited to target infrequent use-cases tailored to a specific entity. %
One limitation of this work is that we relied on OpenAI's API for LLMs. However, we did not observe any significant differences in behavior between the LLMs we experimented with, and we believe that the overall conclusion that template- and LLM-based query generation methods are complementary will remain valid. Another limitation is that the LLM training data can bias the generated query priors, however addressing this is out of the scope of the current work.
Future work includes approaches to mix together the results of the multiple query generation methods, such that the final collection aligns with user behavior; in addition to exploration of the prompt used to query the LLM, use of more advanced prompting techniques (e.g., chain of thought), and LLM fine-tuning. 

\section*{Acknowledgments}

The authors would like to thank %
Manos Tsagkias, %
Lyan Verwimp, %
Russ Web, %
Sameer Badaskar, %
and the anonymous reviewers %
for their comments and feedback.

\section*{Speaker biography}

\noindent \textbf{Sonal Sannigrahi} is a PhD student at Instituto Superior Técnico in Lisbon, Portugal working on multi-modal Natural Language Processing. She previously worked on multilingual representation learning and has published papers at EACL, ACL, amongst others.

\noindent \textbf{Christophe Van Gysel} is a Staff Research Scientist working on the Siri Speech language modeling team at Apple where he works on the boundary between ASR and Search.
Christophe obtained his PhD in Computer Science from the University of Amsterdam in 2017. During his PhD, Christophe worked on neural ranking using representation learning models with a focus on entities and published at WWW, SIGIR, CIKM, WSDM, TOIS, amongst others.

\section*{Company profile}

Apple revolutionised personal technology with the introduction of the Macintosh in 1984. Today, Apple leads the world in innovation with iPhone, iPad, Mac, Apple Watch, and Apple TV. Apple’s five software platforms — iOS, iPadOS, macOS, watchOS, and tvOS — provide seamless experiences across all Apple devices and empower people with breakthrough services including the App Store, Apple Music, Apple Pay, and iCloud. Apple’s more than 100,000 employees are dedicated to making the best products on earth, and to leaving the world better than we found it.

\bibliographystyle{ACM-Reference-Format}
\balance
\bibliography{sigir2024-llm_query_gen}

\end{document}